\documentclass[aps,
physrev,
superscriptaddress,
amsmath,
amssymb,
reprint,
nofootinbib,
floatfix]
{revtex4-1}
\usepackage[dvipdfmx]{graphicx}
\usepackage{mathrsfs}
\usepackage{amsfonts}
\usepackage{times}
\usepackage{amsmath}
\usepackage{leftidx}
\usepackage{tikz}
\usepackage{float}
\usepackage{appendix}
\usepackage{bbold}
\usepackage{braket}
\usepackage{mathtools}
\usepackage{physics}

\newcommand{\kvec}{{\bf k}}

\begin{document}
\title{Variational Quantum Simulation for Periodic Materials}

\author{Nobuyuki Yoshioka}
\email{{\color{black} nyoshioka@ap.t.u-tokyo.ac.jp}}
\affiliation{\color{black} Department of Applied Physics, University of Tokyo,
7-3-1 Hongo, Bunkyo-ku, Tokyo 113-8656, Japan}
\affiliation{Theoretical Quantum Physics Laboratory, RIKEN Cluster for Pioneering Research (CPR), Wako-shi, Saitama 351-0198, Japan}

\author{\color{black} Takeshi Sato}
\affiliation{\color{black}  Graduate School of Engineering, The University of Tokyo, 7-3-1 Hongo, Bunkyo-ku, Tokyo 113-8656, Japan}
\affiliation{\color{black} Photon Science Center, School of Engineering, The University of Tokyo, 7-3-1 Hongo, Bunkyo-ku, Tokyo 113-8656, Japan}
\affiliation{\color{black} Research Institute for Photon Science and Laser Technology, The University of Tokyo, 7-3-1 Hongo, Bunkyo-ku, Tokyo 113-0033, Japan}

\author{Yuya O. Nakagawa}
\affiliation{QunaSys Inc., Aqua Hakusan Building 9F, 1-13-7 Hakusan, Bunkyo, Tokyo 113-0001, Japan}
\author{Yu-ya Ohnishi}
\affiliation{Materials Informatics Initiative, RD Technology \& Digital Transformation Center, JSR Corporation, 3-103-9 Tonomachi, Kawasaki-ku, Kawasaki, Kanagawa, 210-0821, Japan}
\author{Wataru Mizukami}
\email{{\color{black} wataru.mizukami.857@qiqb.osaka-u.ac.jp}}
\affiliation{\color{black} Center for Quantum Information and Quantum Biology,
Osaka University, Toyonaka, Osaka, Japan.}
\affiliation{JST, PRESTO, 4-1-8 Honcho, Kawaguchi, Saitama 332-0012, Japan}
\affiliation{Graduate School of Engineering Science, Osaka University, 1-3 Machikaneyama, Toyonaka, Osaka 560-8531, Japan.}
\date{\today}

\begin{abstract}
We present a quantum-classical hybrid algorithm that simulates electronic structures of periodic systems such as ground states and quasiparticle band structures.
By extending the unitary coupled cluster (UCC) theory to describe crystals in arbitrary dimensions, for a hydrogen chain, we numerically demonstrate  that the UCC ansatz implemented on a quantum circuit can be successfully optimized with a small deviation from the exact diagonalization over the entire range of the potential energy curves.
Furthermore, by using the quantum subspace expansion method, in which we truncate the Hilbert space within the linear response regime from the ground state, the quasiparticle band structure is computed as charged excited states. 
Our work establishes a powerful interface between the rapidly developing quantum technology and modern material science.
\end{abstract}
\maketitle

\emph{Introduction.---}
Achieving decisive ab initio descriptions of electronic properties in solid systems is one of the most significant issues in modern material science.
For weakly correlated systems, the development of the density functional theory (DFT)~\cite{heyd_2003, zhao_2008, tran_2009} and GW approximation~\cite{hybertsen_1985, hybertsen_1986} have realized increasingly accurate numerical simulations.
Wave-function-based techniques have also been studied intensively: time-dependent Hartree-Fock theory~\cite{hirata_1999}, second-order M{\o}ller-Plesset perturbation theory
(MP2)~\cite{sun_1996, hirata_2009} and coupled-cluster (CC) theory with single and
double excitations (CCSD)~\cite{sun_1996, hirata_2001, mcclain_2017}.
More recent reports include CCSD with perturbative triple
excitations (CCSD(T))~\cite{gruneis_201l} and full configuration-interaction (FCI) quantum Monte Carlo method for
periodic solids~\cite{booth_2013}.
Meanwhile, it must be noted that periodic systems contrast sharply with molecular systems, in that one must simulate the thermodynamic limit. 
In general, a large number of particles, or the Brillouin zone sampling, are required to achieve  convergence to the thermodynamic limit. 
The growth of computational resources requirements rapidly exceeds supercomputing capacity, which severely limits the exploration of realistic materials. 
Therefore, algorithms with both favorable scaling and high accuracy beyond the current schemes are indispensable.

The surging development in quantum technology may offer a path to achieving this goal. 
The variational quantum eigensolver (VQE) algorithm and its variants, for instance, enable the simulation of eigenstates of a given Hamiltonian on noisy intermediate-scale quantum (NISQ) devices~\cite{peruzzo_2014}.
Although the rigorous computational speed-up by the VQE-based calculations over classical algorithms still remains elusive (a situation related to the present lack of quantum error correction){\color{black} \cite{Cerezo2021NaturePhysRev}}, many studies have focused on its demonstration in actual quantum devices~\cite{omalley_2016, kandala_2017, kandala_2019, kokail_2019,arute_2020} and its extension to solve a various classes of problems~\cite{nakanishi_2019, colless_2018, higgott_2019, parrish_2019,yoshioka_2019_dvqe, mizukami_2019,Cade2019arXiv, fujii_2020,Endo2020PRR}.
From a quantum chemistry perspective, an intriguing question is whether the NISQ devices become capable of implementing classically intractable wave function ansatz such as the unitary coupled cluster (UCC) ansatz~\cite{Bartlett1989CPL_UCC,Kutzelnigg1991TCA_UCC,Taube2006IJQC_UCC,Yanai2006JCP_CT,Cooper2010JCP_UCC,Harsha2018JCP_UCC}, a variational parametrization of CC wave functions based on unitary transformation.
Although classical computers suffer from an exponential increase in the computational cost, quantum computers naturally simulate such ansatze with only a polynomial number of quantum gates.
For efficient implementation on the NISQ devices, more hardware-friendly and/or sophisticated ansatze have been proposed~\cite{lee_2018, grimsley_2019, matsuzawa_2020}.
However, to the best of our knowledge, existing VQE algorithms and their demonstrations have been mainly performed for small molecules, and periodic systems have not been successfully simulated.

In the present work, we propose and demonstrate a VQE-based framework enables simulations of solid materials at the ab-initio level.
Our work is the first to show that electronic ground states of periodic systems such as the hydrogen chain can be computed accurately even in the strongly correlated regime where the classical gold-standard methods such as CCSD(T) break down.
Furthermore, we present a method to calculate the quasiparticle band structure from the VQE quantum state. 
Our approach is to describe quasiparticle excitations by linear-response-based calculations, i.e., the quantum subspace expansion (QSE)~\cite{mcclean_2017}.
The present work establishes a powerful interface between two major fields, namely the rapidly developing quantum computing technology and modern material science.

\emph{Second quantized ab-initio crystal Hamiltonian.---}
Ab initio fermionic Hamiltonian with periodic boundary conditions is given in the second quantization representation as
\begin{eqnarray}\label{eqn:fermionic_ham}
    \hat{H} &=& \sum_{pq} \sum_{\kvec} t_{pq}^{\kvec} \hat{c}_{p\kvec}^{\dagger} \hat{c}_{q \kvec} \nonumber \\
     &&+ \sum_{pqrs} \sum_{\kvec_p \kvec_q \kvec_r \kvec_s}' v_{pqrs}^{\kvec_p \kvec_q \kvec_r \kvec_s} \hat{c}_{p \kvec_p}^{\dagger} \hat{c}_{q \kvec_q}^{\dagger} \hat{c}_{r \kvec_r} \hat{c}_{s \kvec_s},
\end{eqnarray}
where $\hat{c}_{p \kvec}$ ($\hat{c}_{p \kvec}^{\dagger}$) is the annihilation (creation) operator of the $p$-th Bloch or crystalline orbital (CO) with crystal momentum $\kvec$.  
The complex coefficients $t_{pq}^{\kvec}$ and $v_{pqrs}^{\kvec_p \kvec_q \kvec_r \kvec_s}$ are one- and two-body integrals between COs.
Note that, because of translational symmetry, the two-body term must obey the conservation law 
written as 
\begin{equation}\label{eq:momentum_conserv}
    \kvec_p+\kvec_q -\kvec_r - \kvec_s = {\bf G}, 
\end{equation} 
where ${\bf G}$ is a reciprocal lattice vector of the unit cell. 
Such a requirement is indicated by the primed summation in Eq.~\eqref{eqn:fermionic_ham}.
In the present work, we determine COs from the crystal Hartree--Fock theory with the Gaussian-based atomic orbitals (AOs), for which we employ minimal basis sets (i.e., STO-3G)~\cite{del_1967, andre_1969}. 
Two remarks concerning the computation of integrals are in order. 
First, the divergence corrections for exchange integrals are computed separately. 
Because the ${\bf G}=0$ contribution in the exchange integrals simply shifts the band structure according to their particle number, we initially neglect the divergent term and add the corresponding correction after computing the correlation energy~\cite{pyscf}. 
Second, the Gaussian density fitting technique~\cite{Sun2017JCP_GDF} is used for the two-body coefficients $v_{pqrs}^{\kvec_p \kvec_q \kvec_r \kvec_s}$ to accelerate the integral calculation.

\begin{figure}[t]
    \begin{center}
     \resizebox{0.98\hsize}{!}{\includegraphics{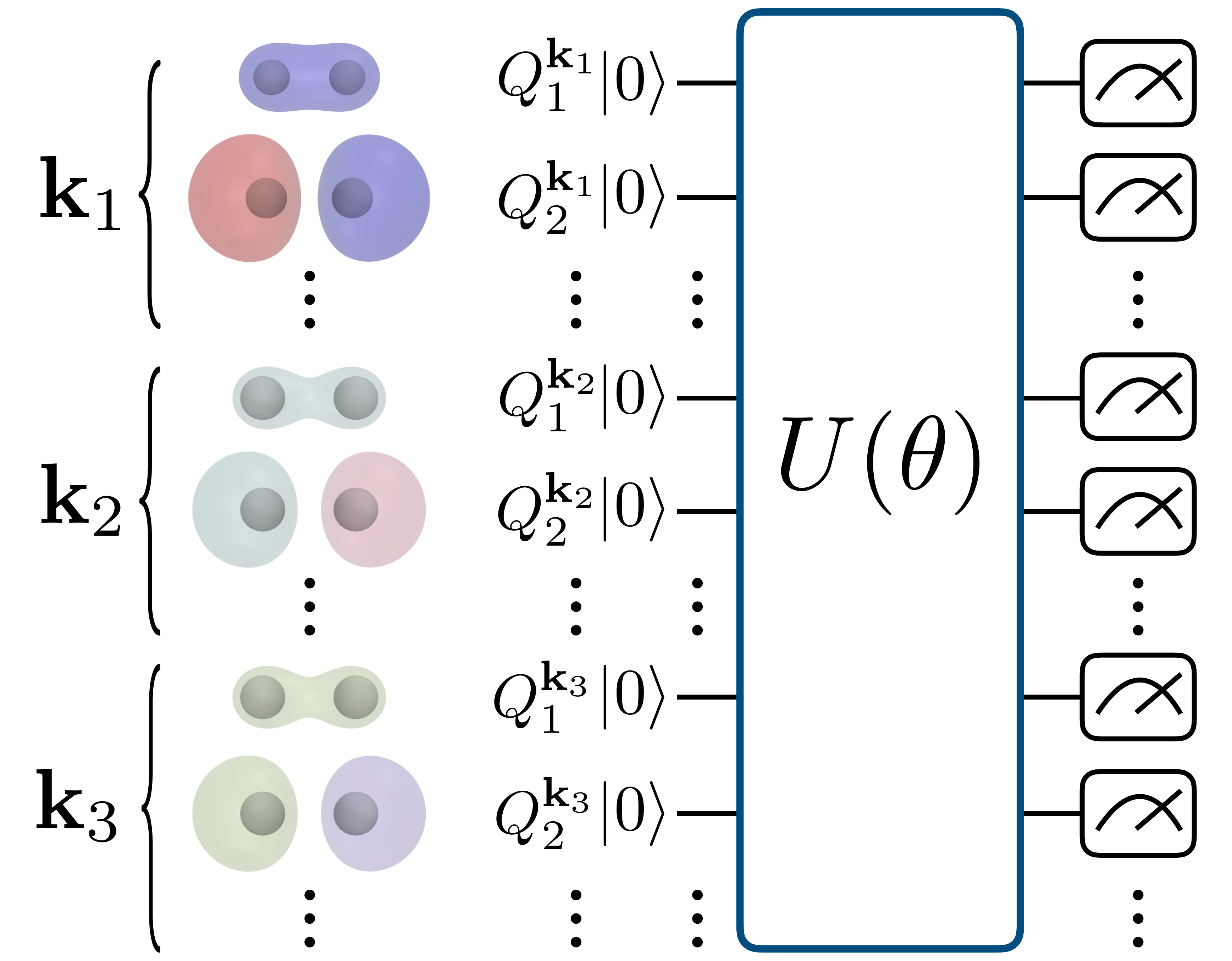}}
   \end{center}
\caption{\label{fig:circuit} Encoding crystalline orbitals into variational quantum circuits. 
The example is for a linear hydrogen chain where each unit cell contains two atoms.
The number of the qubits increases linearly with respect to the number of $k$-points sampled from the Brillouin zone.
}
\end{figure}

To solve the Schr\"{o}dinger equation defined by the Hamiltonian \eqref{eqn:fermionic_ham} on a quantum computer, we map fermionic operators into spin-1/2 operators. 
A widely-known technique is the Jordan--Wigner transformation~\cite{jordan_1928}, which naturally encodes the fermionic anticommutation relation as the parity of the particle number.
Although we adopt the Jordan--Wigner transformation in the present work,
one may also consult on other techniques with improved non-locality~\cite{bravyi_2002}, which may become crucial for  suppressing Pauli measurement error in noisy devices. 
Here, the fermion-qubit mapping algorithm is no different in the case of crystalline systems than in molecular systems. 
However, the number of qubits required in a periodic system could be considerably larger than that in an isolated system; 
if the number of $k$-point samples is set to $N_k$, the number of qubits required is increased by $N_k$ [See Fig.~\ref{fig:circuit} for a graphical description.].

\emph{Variational quantum eigensolver and unitary coupled cluster theory for solids.---}
Once the qubit representation of the crystal Hamiltonian is prepared, 
the ground state wave function and its energy can be calculated on a quantum computer 
using the VQE algorithm. 
That is, one constructs a quantum circuit $\hat{U}(\theta)$ where $\theta$ denotes the circuit parameters, and a trial wave function $\ket{\psi(\theta)} = \hat{U}(\theta) \ket{0}$, where $\ket{0}$ is an input quantum state. 
The ground state is approximated by the variational ansatz $|\psi(\theta^*)\rangle$ whose parameters are taken to minimize the energy function as follows,
\begin{eqnarray}
\theta^* = \mathop{\rm arg~min}\limits_{\theta} E(\theta), \\
E(\theta) := \langle \psi(\theta) | \hat{H} | \psi(\theta) \rangle.
\end{eqnarray} 

Various quantum circuits (i.e., ansatze) have been proposed for the VQE to describe many-body wave functions accurately and compactly. In this study, we choose the unitary coupled cluster singles and doubles (UCCSD) ansatz~\cite{Bartlett1989CPL_UCC,Kutzelnigg1991TCA_UCC,Taube2006IJQC_UCC,Yanai2006JCP_CT,Cooper2010JCP_UCC,Harsha2018JCP_UCC} with one-step Trotter expansion, or the disentangled UCCSD ansatz~\cite{Evangelista2019JCP_dUCC}: 
\begin{equation}\label{eqn:uccsd}
    \ket{\psi} = \left( \prod_{pq} \prod_{\kvec_p \kvec_q} e^{\hat{A}_{pq}^{\kvec_p \kvec_q}} \right)
    \left( 
    \prod_{pqrs} \prod_{\kvec_p \kvec_q \kvec_r \kvec_s} e^{\hat{A}_{pqrs}^{\kvec_p \kvec_q \kvec_r \kvec_s}} 
    \right)
    \ket{0},  
\end{equation}
where 
$\hat{A}_{pq}^{\kvec_p \kvec_q} = a_{pq}^{\kvec_p \kvec_q}\hat{c}_{p \kvec_p}^{\dagger} \hat{c}_{q \kvec_q} \nonumber - c.c.$ 
and 
$\hat{A}_{pqrs}^{\kvec_p \kvec_q \kvec_r \kvec_s} = 
 a_{pqrs}^{\kvec_p \kvec_q \kvec_r \kvec_s}
     \hat{c}_{p \kvec_p}^{\dagger} \hat{c}_{q \kvec_q}^{\dagger} \hat{c}_{r \kvec_r} \hat{c}_{s \kvec_s}
     - c.c.$ 
are cluster operators for single and double excitations, respectively.
The Hartree--Fock state is chosen as the input state $\ket{0}$. 
The coefficients $a$ are the variational parameters and correspond to $\theta$. 
Note that the variational parameters $a$ of the UCCSD ansatz
are in general complex values for a periodic system, 
while those with a standard time-independent molecular Hamiltonian are real because its matrix elements can be chosen to exclude imaginary values.

One of the strengths of the UCC in general is that it enables the symmetry, such as the number of particles or the total spin angular momentum, to be easily introduced into the ansatz (i.e., quantum circuit) because the UCC is based on the fermionic representation.
In addition, the translational symmetry can be straightforwardly implemented into the UCC by limiting $\hat{A}_{pq}^{\kvec_p \kvec_q}$ and $\hat{A}_{pqrs}^{\kvec_p \kvec_q \kvec_r \kvec_s}$ 
to the operators satisfying the crystal momentum conservation law (Eq.~\eqref{eq:momentum_conserv}). 
Nonetheless, in this work, for simplicity, we limit the variational parameters to real numbers and do not impose the momentum conservation condition. 
We tentatively refer to this variant of the UCCSD ansatz as the broken-translational-symmetry UCCSD ansatz with real variables (bUCCSD-Real) here. 
The quantum circuit of the bUCCSD-Real ansatz is virtually identical to UCCSD's quantum circuit for molecular systems. See Ref.~\cite{Whitfield2011MolPhys} for a detailed description of how to realize a UCC ansatz on a quantum circuit.

{\color{black} 
A further strength of UCC is the practicality of parameter optimization.
Physics-based ansatzes such as UCC allow for the preparation of appropriate initial values and removing redundant parameters.
Such features make physics-inspired ansatzes less prone to the vanishing gradient problem in the cost function, called the barren plateau problem \cite{McClean2018NatComm, Cerezo2021NatComm, Holmes2021arXiv}.
However, such ansatzes are unsuitable for NISQs, as the depth of such ansatzes is typically long (e.g., $O(N^4)$ for the UCCSD) and sensitive to quantum hardware noise.
Hardware noise not only makes the results inaccurate but also causes barren plateaus in deep quantum circuits such as UCC, which can make optimization hard \cite{Wang2020arXiv}.
Crucially, the quantum circuit is to be shallow in order to perform the VQE on NISQs, while it is known that a constant depth circuit cannot achieve sufficient expressive power\cite{Bravyi2020PRL}.It is thus vital to strike a good balance between noise and circuit complexity to achieve a practical ansatz.
The variational quantum compiling would be a way to find such a compromise point\cite{Sharma2020NewJPhys}.
Nevertheless, this point is beyond the scope of this paper, and will not be discussed further here. 
}

\begin{figure*}[t]
    \begin{center}
     \resizebox{0.98\hsize}{!}{\includegraphics{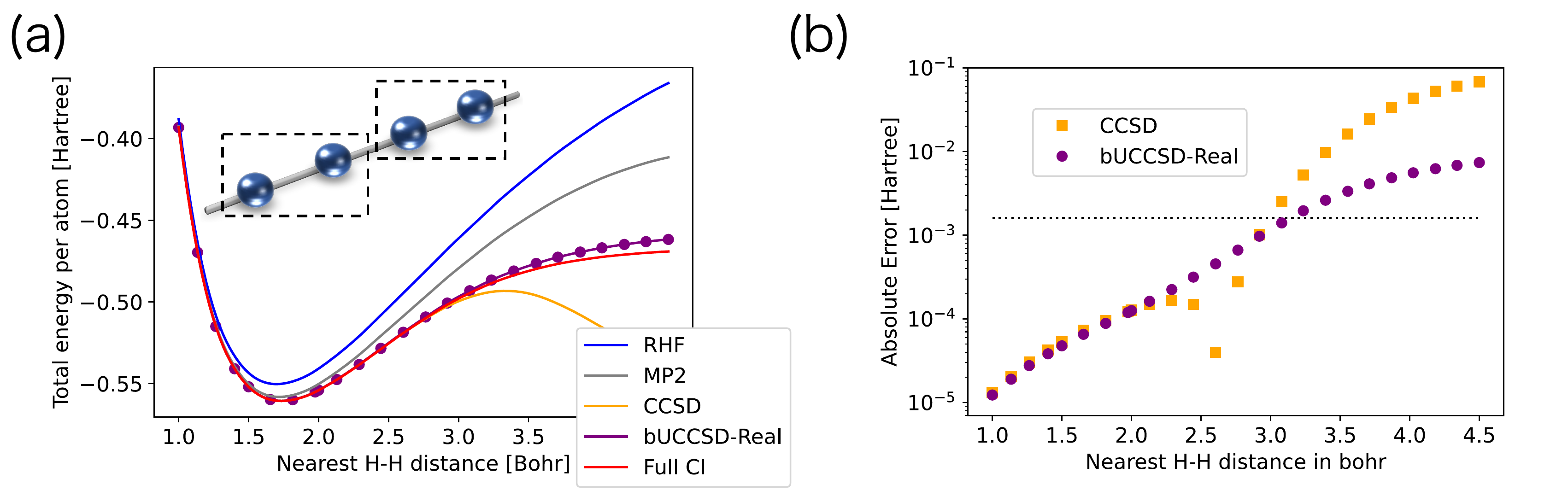}}
   \end{center}
\caption{\label{fig:h2_PEC_uccsd} (a) Potential energy curves of the linear hydrogen chain computed at bUCCSD-Real, CCSD, MP2, RHF, and FCI with the STO-3G basis sets. 
Each unit cell consists of two hydrogen atoms each, and three $k$-points are sampled from  grid. 12-qubits are used for the bUCCSD-Real.
(b) Absolute error from the FCI calculation. In the weakly correlated region, the bUCCSD-Real ansatz are slightly more accurate than CCSD, whose deviation crosses from positive to negative near 2.6 Bohr. The dotted line indicates the chemical accuracy ($1.6\times 10^{-3}$ Hartree).
}
\end{figure*}

\emph{Computing quasiparticle bands from the VQE wave function.---}
Of course, the ground state energy is not the only important and interesting property for solids. 
The force on each nucleus, for example, is often essential in practical calculations, and can be obtained from the energy derivatives. 
Energy derivatives for periodic systems can be calculated in the same manner as those for molecules. 
The analytical energy derivative calculation methods for the VQE quantum state have already been established and applied~\cite{Mitarai2020PRR_derivative, mizukami_2019}. 
Another example is the band structure, which is a property peculiar to solids. 
It is a common and indispensable concept or tool for analyzing the electronic structure of a crystal. 
Band calculations for quantum many-body systems are often performed by simulating  quasiparticle excitations, assuming that the theoretical framework defined in the one-body picture still holds. 
In this context, various classical algorithms, including the GW approximation, have already been proposed to find the quasiparticle bands~\cite{hybertsen_1985, hybertsen_1986}. 
In the present study, we also employ a similar assumption and extend the QSE method to calculate quasiparticle bands from the VQE quantum state. 

The aim of the QSE method is to compute a subset of the entire eigenspectrum in a subspace defined from a reference quantum state $|\psi\rangle$. 
Concretely, we first prepare a set of many-body basis $\ket{\Phi_i} =  \hat{R}_i \ket{\psi}$, where $\ket{\psi}$ is the VQE quantum state and $\hat{R}_i = \hat{c}_{p \kvec_p}^{\dagger} \hat{c}_{q \kvec_q}^{\dagger} \dots \hat{c}_{r \kvec_r} \hat{c}_{s \kvec_s} \dots $ is an excitation operator with a multi-index $i$  specifying a string of annihilation and creation operators. 
Then, we diagonalize a subspace Hamiltonian ${\bf H}^{sub}$ defined in a truncated Hilbert space spanned by $\{ \ket{\Phi_i } \}$. 
The non-orthogonality of such many-body bases requires us to solve the following  generalized eigenvalue problems:
\begin{equation}
    {\bf H}^{sub}{\bf C}={\bf S}^{sub}{\bf C}{\bf E},
\end{equation}
where ${\bf S}^{sub}$ is a metric of the subspace given by the overlap between bases, ${\bf C}$ are eigenvectors, and the diagonal matrix ${\bf E}$ yields eigenenergies.
The matrix elements of the subspace Hamiltonian ${\bf H}^{sub}$ and the metric ${\bf S}^{sub}$ are given by
\begin{align}
    H^{sub}_{ij} &= \langle \Phi_i|H|\Phi_j \rangle = \langle \psi(\theta)|\hat{R}_i^{\dagger} \hat{H} \hat{R}_j|\psi(\theta)\rangle, \\
    S^{sub}_{ij} &= \langle \Phi_i|\Phi_j \rangle= \langle \psi(\theta)|\hat{R}_i^{\dagger} \hat{R}_j |\psi(\theta) \rangle.
\end{align}
These quantities are evaluated as the expectation value of non-Hermitian operators $\hat{R}_i^{\dagger}\hat{H} \hat{R}_j$ and $\hat{R}_i^{\dagger} \hat{R}_j$, which can be realized by measuring  real and imaginary parts separately for instance.

Thus far, the QSE method has been used with particle-number conserving excitation operators, which corresponds to the so-called multi-reference configuration interaction (MRCI) method in the classical algorithms of quantum chemistry~\cite{Werner1988JCP_MRCI,Knowles1988CPL_MRCI}.
The QSE method differs from the MRCI method in that it uses quantum measurements to evaluate the matrix elements. 
Here, we propose a QSE method for calculating quasiparticle bands using many-body bases 
created by ionization or electron-attachment operators, which remove or add one particle from the VQE quantum state $\ket{\psi}$.
The valence band energies are obtained at individual $k$ by performing the QSE using ionization operators $\hat{R}^{IP}_{l \kvec}=\hat{c}_{l \kvec}$ where $l$ runs over the occupied orbitals. 
In contrast, the conduction bands are obtained by using electron-attachment operators $\hat{R}^{EA}_{b \kvec}=\hat{c}^{\dagger}_{b \kvec}$ where $b$ runs over unoccupied orbitals.  
Our method is closely related to a variant of the equation-of-motion coupled cluster (EOM-CC), namely, ionization-potential/electron-attached EOM-CC (IP-EOM-CC, EA-EOM-CC) ~\cite{mcclain_2017}.

\emph{Numerical examples.---}
Now that the theoretical framework is readily provided, we are ready to demonstrate our algorithms in periodic systems.
First, we compute the ground state of the linear hydrogen chain, which is known for its rich physical feature that is still not completely understood despite its simplicity~\cite{stella_2011,lin_2011, mazziotti_2011, yanai_2010, alsaidi_2007, tsuchimochi_2009,sinitskiy_2010,motta_2017}.
The outcome of the electronic interaction varies diversely along the atom separation; the system experiences a metal-insulator transition with a strongly correlated regime in between.
As shown in Fig.~\ref{fig:h2_PEC_uccsd}, the bUCCSD-Real ansatz correctly captures such a complex behaviour.
It is evident from the potential energy curve shown in Fig.~\ref{fig:h2_PEC_uccsd}(a) that strong electronic correlation develops as atoms become separated. 
Therefore, the classical gold-standard CCSD and CCSD(T) methods result in a large deviation from the exact diagonalization, or the FCI; see Fig.~\ref{fig:h2_PEC_uccsd}(b).
In contrast, the bUCCSD-Real ansatz can describe the behaviour of hydrogen atoms much more accurately, owing to the enhanced representability of the variational ansatz.
The bUCCSD-Real ansatz can simulate the weakly correlated region as precise as the CCSD method and suppresses the deviation in the strongly correlated regime.
Considering the fact that the ansatz is not designed to capture the whole Hilbert space with higher-order electronic excitations, we expect that the calculation can be systematically improved by applying more powerful and sophisticated ansatze such as the ADAPT or cluster-Jastrow ansatze~\cite{grimsley_2019, matsuzawa_2020}.

It should be noted that, although the result by the bUCCSD-Real ansatz is presented in the current work, it may be desirable to employ an ansatz with complex variables {\color{black} (such an example is shown in Appendix ~\ref{sec:uccsd_complex})}.
Extending real variables to complex variables results in effectively doubling the number of variables. 
Nonetheless, the disadvantages of extending to complex variables are presumably compensated by using the momentum conservation law (Eq.~\eqref{eq:momentum_conserv}); the translational symmetry leads to a considerable reduction in the number of parameters, especially when many $k$-points must be considered, such as in the three-dimensional systems.

\begin{figure}[t]
    \begin{center}
     \resizebox{0.98\hsize}{!}{\includegraphics{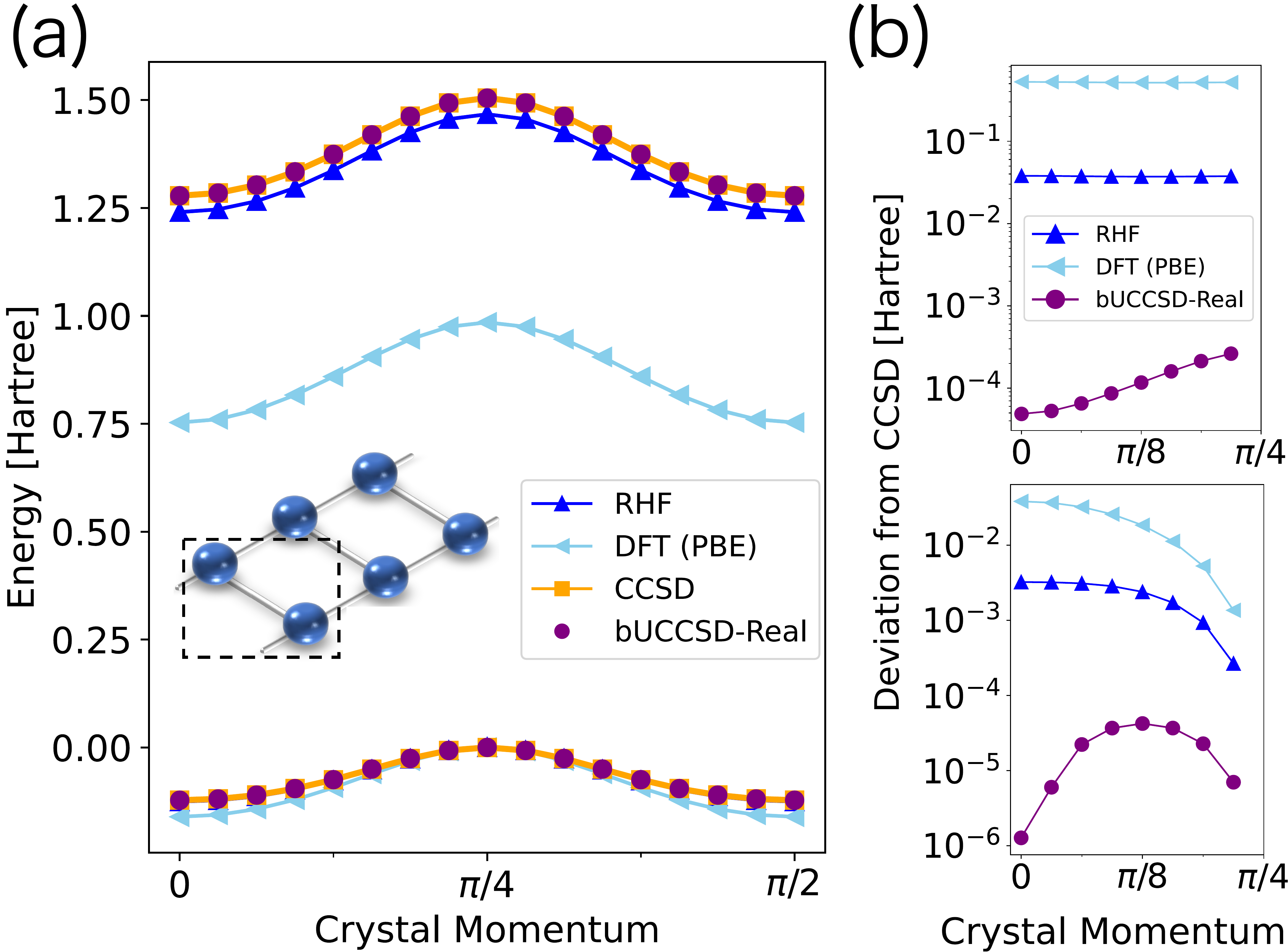}}
   \end{center}
\caption{\label{fig:h2_dimer_chain_band} (a) Band structure of the hydrogen dimer chain computed at bUCCSD-Real, CCSD, and RHF with the STO-3G basis sets. The energy is shifted so that the highest energy of the occupied band is zero. Two bands are well separated by a gap owing to the coherent spin singlet formation.
(b) Absolute deviation of the electron affinity (upper panel) and ionization potential (lower panel) from the equation-of-motion CCSD calculation. 
A unit cell considered in the calculation consists of a pair of hydrogen atoms that are 1.2 Bohr apart from each other, and the distance between dimers is taken as 4 Bohr. Two $k$-points are sampled from a uniform grid. 8-qubits are used for the bUCCSD-Real.
}
\end{figure}

Next, we turn to the band-structure calculation of the hydrogen dimer chain.
Such two-leg ladder systems are of strong interest from both the theoretical and experimental aspects, because synthesized compounds on ladder structures may show exotic phenomena such as the unconventional superconductivity and spin-liquid behavior~\cite{dagotto_1996}. 
In particular, the half-filled Hubbard model on a two-leg ladder is gapped by both charge and spin excitations, as opposed to that on the linear chain. 
Such a state with spin singlets on each rung has been found to evolve into a superfluid phase by additional spin exchange interaction between rungs~\cite{tsuchiizu_2002}.
Here, we take a large distance between hydrogen dimers so that the system is described by the coherent spin singlet state.
The quasiparticle spectrum of the system is obtained by the ionized/electron-attached QSE method introduced previously. 
As can be seen from Fig.~\ref{fig:h2_dimer_chain_band}, both the highest occupied and lowest unoccupied bands are simulated precisely.
In particular, the direct band gap estimated at crystal momentum $kL=\pi/4$ ($L$: unit cell length) is 1.5047 Hartree, which is consistent with the EOM-CCSD calculation with an error less than $3\times 10^{-4}$ Hartree.

\emph{Summary and outlook.---}
We have presented a framework for simulating the electronic structures of solids using NISQ devices at the ab initio level.
The numerical results demonstrate that our VQE-based algorithm simulates the hydrogen chain well not only in the weakly-correlated electronic structures but also for the strongly correlated regimes.
Furthermore, we have shown that the quasiparticle band structure can be computed by applying the QSE method, which diagonalizes the Hamiltonian in a truncated space described by the linear response, to charged excited states. 
The use of ionization/electron-attachment operators yields a substantial improvement in the measurement cost that scales quadratically with respect to the qubit count, as opposed to the quartic (or higher) scaling required in the standard QSE method which employs particle-number-conserved excitation operators. 

Our VQE-based framework is expected to provide an approach for  investigating otherwise intractable systems. 
In addition to the insulating low-dimensional materials calculated in the present work,
real solid surface systems and strongly correlated materials are core targets that should be investigated once the quantum computers become sufficiently mature.
To address target materials having a tangible impact not only for scientific knowledge but also for industrial applications, it would be necessary to develop a qubit reduction technique that explicitly makes use of the symmetry.


\emph{Acknowledgements.---}
This work was supported by MEXT Quantum Leap Flagship Program (MEXT Q-LEAP) Grant Number JPMXS0118067394 {\color{black} and JPMXS0120319794}.
WM wishes to thank JSPS KAKENHI No.\ 18K14181 and JST PRESTO No.\ JPMJPR191A.
Some calculations were performed using the computational facilities in the Institute of Solid State Physics at the University of Tokyo, RIKEN,
and in Research Institute for Information Technology (RIIT) at Kyushu University, Japan.
NY was supported by the Japan Science and Technology Agency (JST) (via the Q-LEAP program).
{\color{black} We also acknowledge support from JST COI-NEXT program.}
We thank Dr. Kosuke Mitarai for fruitful discussions. 
Numerical calculations were performed using OpenFermion~\cite{openfermion}, PySCF (v1.7.1)~\cite{pyscf}, and  Qulacs~\cite{qulacs}.

\emph{Note added.---}
Shortly after completion of this work, we became aware of two independent works~\cite{liu_2020, manrique_2020} that have been carried out in parallel.


\appendix

\section{Numerical results for momentum conservations and fidelities \label{sec:momentum}}
\begin{figure}
\includegraphics[width=0.49\textwidth]{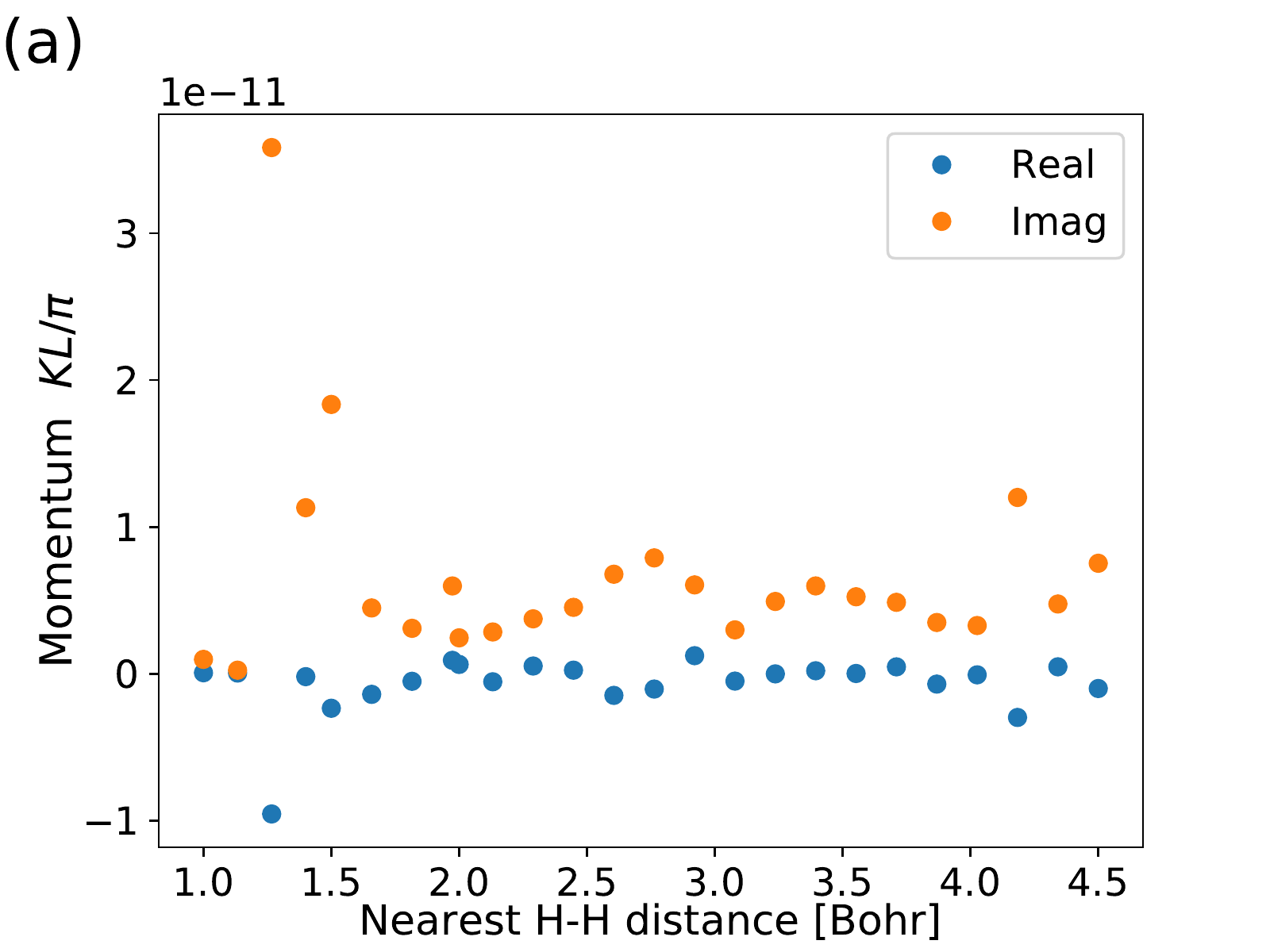}
\includegraphics[width=0.49\textwidth]{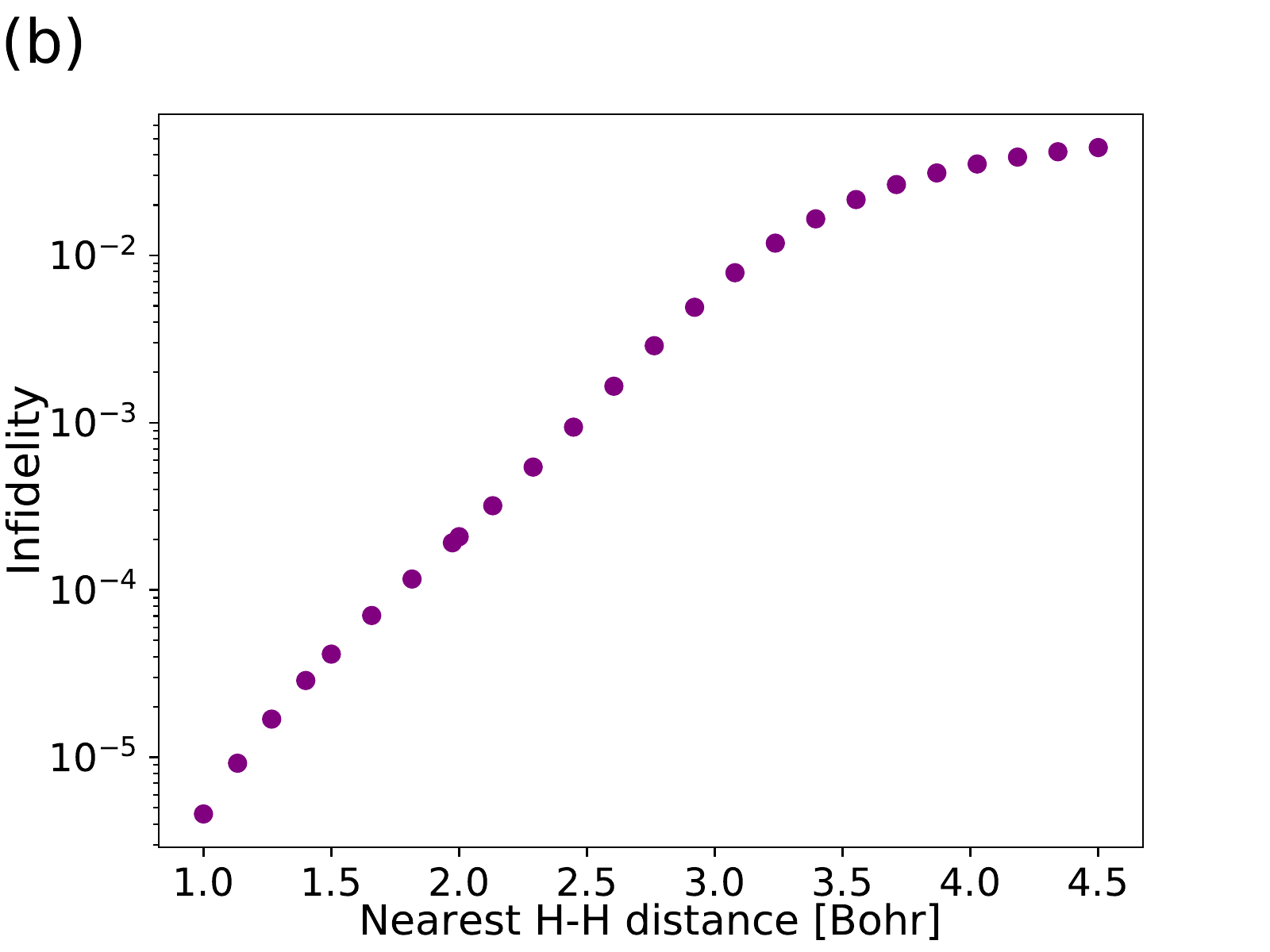}
\caption{\label{appfig:h2 chain}
(a) Crystal momentum of the optimized bUCCSD-Real wavefunction whose energy is shown in Fig.~\ref{fig:h2_PEC_uccsd}(a) of the main text. A unit cell with the length $L$ consists of two hydrogen atoms each, and three $k$-points are sampled from a uniform grid. 12-qubits are used for the bUCCSD-Real.
Crystal momentum $K$ is determined by the equation $e^{iKL} = \Braket{\psi_{\mathrm{bUCCSD-R}} | \hat{T}_{L} | \psi_{\mathrm{bUCCSD-R}}}$, where $\hat{T}_L$ is the translation operator of the length $L$. 
(b) Infidelity between the optimized bUCCSD-Real wavefunction and the exact wavefunction obtained by the FCI calculation. 
}
\end{figure}

As explained in the main text, we adopt the broken-translational-symmetry UCCSD ansatz with real variables (bUCCSD-Real) for demonstration of the variational quantum simulation for periodic materials. 
The total crystal momentum can be different from the input state of the ansatz, i.e., the Hartree-Fock state, in this ansatz.
We numerically check the total crystal momentum $K$ of the optimized wave function $\ket{\psi_{\mathrm{bUCCSD-R}}}$ for the linear hydrogen chain investigated in Fig.~\ref{fig:h2_PEC_uccsd} of the main text.
We compute the expectation value of the translation operator $\hat{T}_L$ defined for the unit cell length $L$ as $\Braket{\psi_{\mathrm{bUCCSD-R}} | \hat{T}_{L} | \psi_{\mathrm{bUCCSD-R}}}$, 
and this expectation value is identified with $e^{iKL}$.
The result is shown in Fig.~\ref{appfig:h2 chain}(a).
As is expected from the uniform geometry of the chain, the total crystal momentum $K$ is apparently zero within the numerical error caused by slight imperfectness of the optimization of the wavefunction in the VQE: the mean of real and imaginary parts of $KL/\pi$ for all data points are below $10^{-11}$.

In addition, we present infidelity between  $\ket{\psi_{\mathrm{bUCCSD-R}}}$ and the exact wavefunction computed by the full configuration-interaction (FCI) $\ket{\psi_{\mathrm{FCI}}}$, namely $1-|\Braket{\psi_{\mathrm{bUCCSD-R}}|\psi_{\mathrm{FCI}}}|$, in Fig.~\ref{appfig:h2 chain}(b).
The infidelity is small and varies with the hydrogen-hydrogen distance of the chain in a similar way to the deviation from the exact energy (Fig.~\ref{fig:h2_PEC_uccsd}(b)).
This result also indicates that our choice of the ansatz properly reproduces the exact wavefunction of the system.


\section{Details of quantum circuits used in numerical experiments \label{sec:uccsd_circuit}}
We used the software Qulacs to simulate quantum circuits employed in our numerical calculations of the VQE with the bUCCSD-Real ansatz. The VQE parameters were optimized using the BFGS algorithm implemented in the SciPy package. The number of qubits employed were 12-qubits and 8-qubits  for the linear hydrogen chain and the hydrogen dimer chain, respectively. The depth of the bUCCSD-Real quantum circuits were 964 and 192 for the linear chain and the dimer chain; the number of the VQE parameters $\theta$ were 1044 and 208.  Note that the quantum circuits used in this study were not prepared as a directly transferable one to any particular quantum computer architecture that currently exists.  First, the quantum circuits were not prepared with a specific qubit connectivity in mind. Second, the multi-qubit gates appearing in UCC ansatzes (including the bUCCSD-Real) were not decomposed into 2-qubit or 1-qubit gates. Also, we performed a perfect simulation without considering any noise in this study. Nonetheless, these differences between the existing quantum computers do not affect numerical simulation results, as long as the absence of noise is assumed.

{\color{black}
\section{Comparison of real- and complex-variable UCC ansatzes \label{sec:uccsd_complex}}

\begin{figure}
\includegraphics[width=0.49\textwidth]{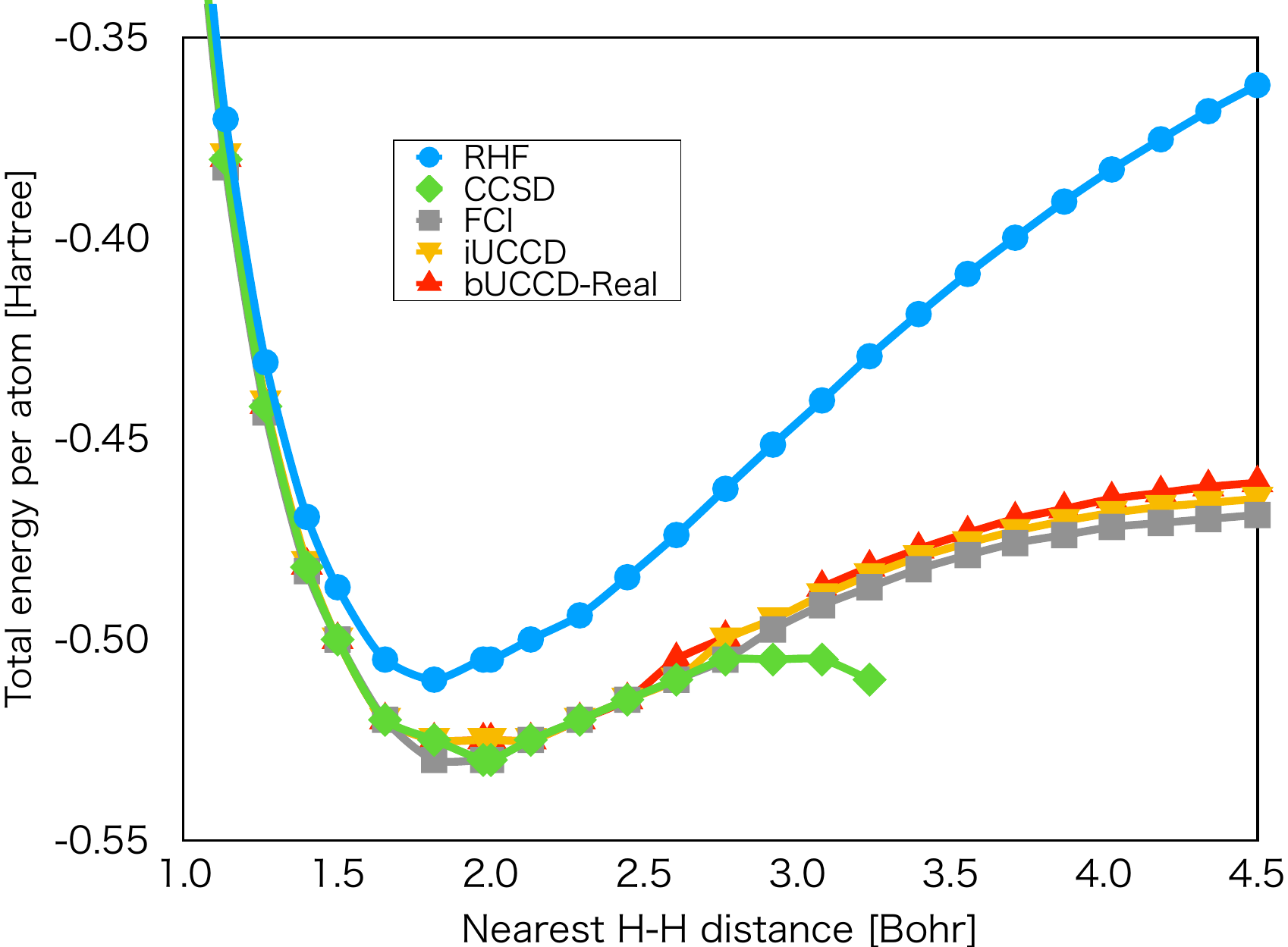}
\caption{\label{appfig:h2 chain_four-k-point}
 \color{black} Potential energy curves of the linear hydrogen chain computed with four k-points at bUCCD-Real, complex variable UCCD (iUCCD) with translational symmetry, RHF, CCSD and FCI with the STO-3G basis sets. Each unit cell consists of two hydrogen atoms each, and k-points are sampled from uniform grid.}
\end{figure} 

%

The UCC without imaginary amplitudes (UCC-Real) reproduced the FCI results well for the systems studied in the main text. However, as mentioned in the main text, this is not always the case. Indeed there are systems where the UCC-Real has a lower descriptive power because of the lack of complex coefficients. A hydrogen chain with four k-points is such an example. A comparison of UCC-Real with the UCC with complex amplitudes (iUCC) was performed for this system. To reduce computational costs, we reduced the number of parameters by using UCC with double excitations (UCCD), where singles excitations are omitted. The UCCD with complex coefficients (iUCCD) employed the translational symmetry for further reduction in the number of circuit parameters. Fig S2. shows the potential energy curves (PECs) of the hydrogen chain calculated by iUCCD and the real-variable UCCD with broken translational symmetry (bUCCD-Real). The PEC of iUCCD shows a good agreement with FCI, whereas that of bUCCD-Real has larger deviation from FCI in the strongly-correlated regime. At R= 4.5 Bohr, the errors with respect to the FCI are 4.8 kcal/mol and 10.3 kcal/mol for iUCCD and bUCCD-Real, respectively. In the weakly-correlated regime, the difference between iUCCD and bUCCD-Real becomes smaller. The deviations from FCI are 2.78 kcal/mol and 3.28 kcal/mol for iUCCD and bUCCD-Real, respectively, at R= 2.0 Bohr.
}

%

\end{document}